\documentclass[12pt]{article}
\usepackage{amsfonts,amsmath,amssymb}
\usepackage{theorem}
\usepackage{graphics,epsfig}
\numberwithin{equation}{section}

\theorembodyfont{\rmfamily}

\def\cA{{\cal A}}

          \def\cN{{\cal N}}          
                    
                    \def\cU{{\cal U}}

 \def\mS{{\mathfrak S}}
\def\mW{{\mathfrak W}}

\newcommand{\CC}{{\mathbb C}}
\newcommand{\II}{{\mathbb I}}
\newcommand{\NN}{{\mathbb N}}

\newcommand{\RR}{{\mathbb R}}
\def\lddots{\mathinner{\mkern1mu\raise1pt\hbox{.}\mkern2mu
\raise4pt\hbox{.}\mkern2mu\raise7pt\vbox{\kern7pt\hbox{.}}\mkern1mu}}

\newcommand{\ie}{{\it i.e.}\ }

\def\qmbox#1{\qquad\mbox{#1}\quad}

\begin{document}

\pagestyle{empty} \setcounter{page}{0}


\strut\hfill{}

\vspace{0.5cm}

\begin{center}

{\LARGE \textsf{Exact energy spectrum for models\\
 with equally spaced point potentials \\[5mm]
}}

\vspace{10mm}

{\large V.Caudrelier$~^\dagger$ and N.Cramp\'e$~^\ddagger$}

\vspace{10mm}

\emph{Department of Mathematics\\
 University of
York\\
 Heslington York\\
  YO10 5DD, United Kingdom}

\end{center}

\vfill \vfill

\begin{abstract}
We describe a non-perturbative method for computing the energy
band structures of one-dimensional models with general point
potentials sitting at equally spaced sites. This is done thanks to
a Bethe ansatz approach and the method is applicable even when
periodicity is broken, that is when Bloch's theorem is not valid
any more. We derive the general equation governing the energy
spectrum and illustrate its use in various situations. In
particular, we get exact results for boundary effects. We also
study non-perturbatively the effects of impurities in such
systems. Finally, we discuss the possibility of including
interactions between the particles of these systems.
\end{abstract}

\begin{center}
\textit{In memory of Daniel Arnaudon}
\end{center}

\vfill PACS numbers: 03.65.Fd, 71.20.-b, 71.55.-i

\vspace{1cm}

Keywords: Energy band structures, Impurities, Non-perturbative
methods, Bethe ansatz. \vfill

\rightline{cond-mat/0511619}\vspace{0.5cm}

$^\dagger$ vc502@york.ac.uk

$^\ddagger$ nc501@york.ac.uk

\baselineskip=16pt

\newpage
\pagestyle{plain}

\section*{Introduction}

The study of the so-called point or contact interactions covers an
impressively large number of areas ranging from pure mathematics,
where the main goal consists in defining and understanding them
rigorously, to applied physics, where either they serve as good
approximations for physical situations or they are created to
control the behaviour of certain quantum systems. One particular
case of such contact interactions but certainly the most
well-known is the $\delta$ potential which played a crucial role
in two specific areas of special interest for us in this paper:
solid state physics and exactly and integrable models (in one
spatial dimension). Indeed, in the first context, it was used by
Kronig and Penney \cite{KP} to model a free electron moving in a
crystal lattice and is now a standard of textbooks in solid state
physics. As for the second context, it has become one of the
cornerstones of exactly solvable and integrable systems through
the works of McGuire, Lieb and Liniger and Yang \cite{MG,LL,Y}
concerning a system of identical particles interacting through the
$\delta$ potential. In both cases, the corresponding quantum
mechanical problem can actually be completely solved by taking a
particular form of the eigenfunctions and one obtains the
information on the energy spectrum.

In the Kronig-Penney model, the periodicity of the potential and
of the boundary conditions on the wavefunction allows to use the
famous Bloch's theorem \cite{Bloch} which actually fixes the form
of the eigenfunctions. In the Lieb-Liniger model, one assumes that
the wavefunctions can be expanded on plane waves with appropriate
coefficients that must be found. These approaches combined with
the various conditions imposed on the wavefunctions of the problem
lead to the allowed energy states of the models. Depending on the
context, this gives rise to the famous energy band structures or
to the famous Bethe ansatz equations\cite{Bethe}.

Several generalizations of the two previous problems have been
considered over the past decades. For instance, in \cite{SMMC,EG},
the Kronig-Penney model is extended to the case where the $\delta$
potential is replaced by a more general point potential. In
\cite{Gaudin}, boundary Bethe ansatz equations were derived by
putting bosons with $\delta$ interactions in a box while in
\cite{bart}, impurity Bethe ansatz equations appeared by including
a general external contact potential in a system of particles with
$\delta$ interactions.

With the advent of nanostructures and the ever increasing need for
controlling quantum devices, the standard assumption of
periodicity is no longer accurate enough. But then, Bloch's
theorem no longer applies and, to the best of our knowledge, only
approximate methods are used. In this paper, we mainly address the
above problem and propose a general method to investigate the
energy spectrum in one-dimensional models with equally spaced but
otherwise arbitrary point potentials. Let us note that the
breaking of periodicity can have various origins. For example, one
can simply imagine that the periodic boundary conditions are
replaced by Dirichlet or Neumann boundary conditions. Or, one
could replace the $\delta$ potential at one or more sites by a
different point potential. We will see that the Bethe ansatz
approach brings an elegant and more general alternative to Bloch's
theorem in order to treat such problems. In all cases, we reduce
the problem to finding roots of polynomials and solving Bethe
ansatz equations. We also discuss the possibility of including
interactions between bosons moving in such general potentials.

The paper is organised as follows. In section \ref{gen_pb}, we
discuss the general setup for treating free particles in an
arbitrary equally spaced, external point potential. Then, in
section \ref{period_pot}, we validate our method in the context of
periodic potentials by deriving a general result which encompasses
previous results in the literature and in particular the
well-known features of the Kronig-Penney model. Section \ref{box}
illustrates the use of our method in a context where periodicity
is broken by the boundary conditions. This provides
non-perturbative results for the boundary effects. This is further
illustrated in Section \ref{impurity} where we introduce
impurities at one or several sites of the potential. Again the
results are non-perturbative and this allows for a study of the
effects of impurities with arbitrary strength. Finally, in Section
\ref{interaction}, we discuss the extension of our method to the
case of bosons with $\delta$ interactions moving in a general
point potential.

\section{The general problem}\label{gen_pb}

We study a one-dimensional system of free particles moving in the
interval $[-ML,ML]$, $M\geqslant 1$ being an integer and $L$ a
length scale, with an external point potential sitting at each
site $x^0_j=(M-2j+1)L$, $j=1,\ldots,M$. It is sufficient to
consider the one-particle Hamiltonian which takes the following
form\footnote{In this paper, we use units such that $\hbar=1=2m$.}
\begin{eqnarray}
\label{ham} H=-\frac{d^2}{dx^2} + \sum_{j=1}^M
v_j\big(x-x^0_j\big)\,,
\end{eqnarray}
where, for $j=1,\ldots,M$,
\begin{eqnarray}
\label{litv} v_j(x)= c_j~\delta(x)
+4\lambda_j~\frac{d}{dx}\delta(x)\frac{d}{dx}
+(2\gamma_j+i\eta_j)~\frac{d}{dx}\delta(x)
-(2\gamma_j-i\eta_j)~\delta(x)\frac{d}{dx}\,,
\end{eqnarray}
and $c_j$, $\lambda_j$, $\gamma_j$, $\eta_j\in \RR$. These very
singular point potentials find their origin in the self-adjoint
extensions of the free Hamiltonian $-\frac{d^2}{dx^2}$ on the
space $C^\infty_0(\RR\setminus\{0\})$ of $C^\infty$ functions with
compact support separated from the origin. Written in this way,
they are in fact quite formal but it is known that they are
equivalent to imposing appropriate boundary conditions on the
wavefunction at each site $x_j^0$ parametrized by $U(2)$ matrices
(see for example \cite{ADK} and references therein).

\begin{figure}[htp]
\begin{center}
\epsfig{file=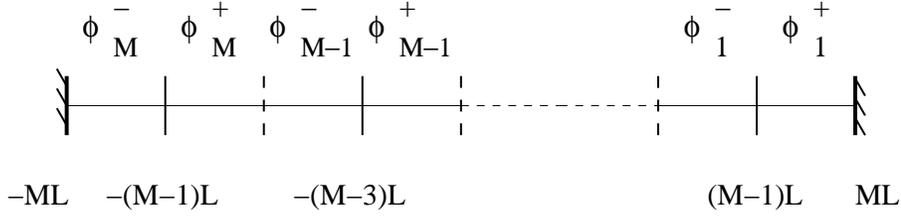,width=12cm}
\caption{\label{fig-explanation}The nontrivial potential lies at
the points $-(M-1)L, -(M-3)L,\ldots, (M-1)L$ (solid lines). The
dashed lines correspond to a trivial potential used for
convenience.}
\end{center}
\end{figure}

In each region $R_j^\pm$ : $(M-2j+\frac{1\pm
1}{2})L<x<(M-2j+\frac{3\pm 1}{2})L$, the particle is actually free
and we denote the wavefunction by $\phi^\pm_j(x)$ (see Figure
\ref{fig-explanation}). Now the idea is to impose all the boundary
conditions (those corresponding to the potential at each $x_j^0$
and those at the ends of the interval) in a compact form by
extending the approach of \cite{bart}. To do so, we collect all
the pieces of the wavefunction into a single $2M$-component vector
defined for $x\in ]ML-L,ML[$
\begin{eqnarray}
\label{Phi}
\Phi(x)=\sum_{j=1}^M \phi^+_j\big(x+2L(j-1)\big)
e_{j}\otimes \hat{e}_{+} +\phi^-_j\big(-x+2L(M-j)\big)
e_{j}\otimes \hat{e}_{-}\,,
\end{eqnarray}
where $\{e_j|1\leqslant j \leqslant M\}$ is the canonical basis of
$\CC^{M}$ and $\{\hat{e}_{+},\hat{e}_{-}\}$ is that of $\CC^2$.
We are now ready to formulate the problem. As we said, this is
just the free problem in each region $R_j^\pm$. In terms of $\Phi$
this reads,
\begin{eqnarray}
\label{free} -\frac{d^2}{dx^2}~\Phi(x)=E~\Phi(x)~~,~~ML-L<x<ML\,.
\end{eqnarray}
Then, following for example \cite{ftc}, the $M$ conditions
corresponding to the general point potential are written as
follows
\begin{eqnarray}
\label{bound_condition}
(\cU^+-\II)~\Phi(x)+i(\cU^++\II)~\frac{d}{dx}\Phi(x)=0 \qmbox{for}
x\rightarrow ML-L\,,
\end{eqnarray}
where $\II$ is the $2M\times 2M$ identity matrix and
\begin{eqnarray}
\label{U} \cU^+=\left(\begin{array}{ccc}
                 U_1 &  &    \\
                    &  \ddots &   \\
                    &  & U_{M}  \\
                 \end{array}\right)\,.
\end{eqnarray}
Each submatrix $U_j$ of $\cU^+$ corresponds to the potential $v_j$
in (\ref{ham}). It is a $U(2)$ matrix parametrized by
\begin{equation}
\label{def_U} U_j=e^{i\xi_j}\left(\begin{array}{cc}
                        \mu_j & \nu_j \\
                         -\nu_j^* & \mu_j^* \\
                       \end{array}\right)~~,~~\xi_j\in[0,\pi),~\mu_j,\nu_j\in\CC~~\text{such
                       that}~~|\mu_j|^2+|\nu_j|^2=1 \,
                       .
\end{equation}
where the symbol $^*$ stands for complex conjugation. The
parameters in the matrix $U_j$ are related to those of the
potential (\ref{litv}). For a discussion of the various
parametrizations of point potentials see e.g. \cite{EG} and
references therein. Finally, the boundary conditions at the ends
of the interval can be written in the same form as before
\begin{eqnarray}
\label{bound_condition2}
(\cU^--\II)~\Phi(x)+i(\cU^-+\II)~\frac{d}{dx}\Phi(x)=0 \qmbox{for}
x\rightarrow ML
\end{eqnarray}
where
\begin{eqnarray}
\cU^-=\left(\begin{array}{ccccccc}
                 e^{i\xi_q}\mu_q &  &&&  && e^{i\xi_q}\nu_q \\
                    &  0&1 &  &  \\
                    & 1 &0 \\
                    & & & \ddots &  \\
                    &  &&  & 0&1 \\
                    &  &&  & 1&0 \\
                    -e^{i\xi_q}{\nu}^*_q&&&&&&e^{i\xi_q}\mu^*_q
                 \end{array}\right)\,.
\end{eqnarray}
The four coefficients at the corners of $\cU^-$ also form a $U(2)$
matrix
\begin{equation}
\label{def_Uq} U_q=e^{i\xi_q}\left(\begin{array}{cc}
                        \mu_q & \nu_q \\
                         -\nu_q^* & \mu_q^* \\
                       \end{array}\right)\,,
\end{equation}
with the same constraints as in (\ref{def_U}). They represent very
general boundary conditions encompassing the usual periodic,
anti-periodic and box boundary conditions. These coefficients
encode the behaviour of the wavefunction on the boundaries $x=-ML$
and $x=ML$ (see Fig. \ref{fig-explanation}). The submatrices
$\left(\begin{array}{cc}
                        0 & 1 \\
                         1 & 0 \\
                       \end{array}\right)$
of $\cU^-$ correspond to the dashed lines in Figure
\ref{fig-explanation}. They are introduced for mere convenience in
our approach.

It is now obvious that one cannot use Bloch's theorem to solve
this problem as explained in the introduction. The potential,
albeit sitting on equally spaced sites, is certainly not periodic
and we impose quite general boundary conditions.

Instead, we formulate a \textit{Bethe ansatz} \cite{Bethe} for
$\Phi$
\begin{eqnarray}
\label{ansatz} \Phi(x)= e^{ikx}\cA_I~+~e^{-ikx}\cA_R \qmbox{where}
\begin{cases}
\displaystyle \cA_I=\sum_{j=1}^{M}\sum_{\epsilon=\pm}
A_{I,j}^\epsilon
~e_{j}\otimes \hat{e}_{\epsilon}\\
\displaystyle \cA_R=\sum_{j=1}^{M}\sum_{\epsilon=\pm}
A_{R,j}^\epsilon ~e_{j} \otimes \hat{e}_{\epsilon}
\end{cases}
\end{eqnarray}
In this form, $\Phi(x)$ is automatically solution of (\ref{free})
with
\begin{equation}
E=k^2\;.
\end{equation}
Inserting the ansatz in (\ref{bound_condition}) and
(\ref{bound_condition2}), we find that $\Phi$ is the eigenfunction
we look for if and only if the $4M$ amplitudes
$A_{I,j}^{\epsilon}$ and $A_{R,j}^{\epsilon 1}$ satisfy the
following relations
\begin{eqnarray}
\cA_R&=&e^{2ik(M-1)L}Z^+(-k)\cA_I\,,\\
\cA_R&=&e^{2ikML}Z^-(-k)\cA_I\,,
\end{eqnarray}
where
\begin{eqnarray}
\label{defZ} Z^\pm(k)=-[\cU^\pm-\II-k
(\cU^\pm+\II)]^{-1}~[\cU^\pm-\II+k (\cU^\pm+\II)]\,.
\end{eqnarray}
These $2M\times 2M$ matrices characterize the type of potential
and boundary conditions one is considering. Requiring a non
trivial solution for the wavefunction, one finds the following
equation
\begin{eqnarray}
\label{bethe} \det\big(Z^+(-k)- Z^-(-k)e^{2ikL}\big)=0\;.
\end{eqnarray}
This type of equation is usually called Bethe ansatz equations. To
our knowledge, it is the first time that the Bethe equations has
been established in this context. They impose constraints on the
allowed values of $k$ which incorporate the effect of the
potential and of the boundary conditions. Solving in $k$ as a
function of the parameters controlling the potential and the
boundary conditions allows one to determine the energy spectrum
and to study how one can modify it by tuning these parameters. As
we will see, this has consequences on the energy band structure of
the associated model.

The above Bethe ansatz equations together with the method to get
them for the very general model we are considering constitute the
main result of this paper. They replace Bloch's theorem when
periodicity is broken and provide a non-perturbative means to get
the energy spectrum in such situations.

\section{Periodic potential}\label{period_pot}

Before exploring some applications of our method to more
complicated cases in the following, we first show in this section
that our approach consistently reproduce well-known results for
periodic potentials and in particular for the Kronig-Penney model.
We also obtain information on the bound states which are less
studied. Finally, we compute the spectrum for an asymmetric
potential which, to our knowledge, is not studied in the
literature.

\subsection{General case}

In our language, a periodic potential with periodic boundary
conditions is obtained by setting
\begin{eqnarray}
\label{perio_condition1}
U_1&=&\dots=U_M=e^{i\xi}\left(\begin{array}{cc}
                        \mu & \nu \\
                         -\nu^* & \mu^* \\
                       \end{array}\right)\equiv U\\
\label{perio_condition2} \qmbox{and} U_q&=& \left(
\begin{array}{cc}
  0 & 1 \\
  1 & 0 \\
\end{array}\right)\,.
\end{eqnarray}
In this case, the Bethe equations (\ref{bethe}) reduce to
\begin{eqnarray}
\left|\begin{array}{ccccccccc}
R^+&T^+&&&&&&-e^{2ikL}\\
T^-&R^-&-e^{2ikL}&&\\
&-e^{2ikL}&R^+&T^+\\
&&T^-&R^-&-e^{2ikL}&&\\
&&&&\ddots\\
&&&&T^-&R^-&-e^{2ikL}&\\
&&&&&-e^{2ikL}&R^+&T^+\\
-e^{2ikL}&&&&&&T^-&R^-
\end{array}\right|=0
\end{eqnarray}
where, for $\mu=\mu_R+i\mu_I$ and $\nu=\nu_R+i\nu_I$, ($\mu_R$,
$\mu_I$, $\nu_R$, $\nu_I$$\in\RR$)\footnote{In comparison with the
notation of \cite{bart}, $R^+=R^+(-k)$, $T^+=T^+(-k)$,
$R^-=R^-(k)$ and $T^-=T^-(k)$}
\begin{eqnarray}
\label{RTRT1} R^+ = \frac{(\cos\xi+\mu_R)k^2 - 2i\mu_Ik
-\cos\xi+\mu_R} {(\cos\xi+\mu_R)k^2 + 2ik\sin\xi+\cos\xi-\mu_R} \,
, ~~ T^+ = \frac{-2\nu k}{(\cos\xi+\mu_R)k^2 + 2ik\sin\xi
+\cos\xi-\mu_R}\,,~~
\\
\label{RTRT2} R^- = \frac{(\cos\xi+\mu_R)k^2 + 2i\mu_Ik
-\cos\xi+\mu_R} {(\cos\xi+\mu_R)k^2 + 2ik\sin\xi  +\cos\xi-\mu_R}
\, , ~~ T^- = \frac{2\nu^* k}{(\cos\xi+\mu_R)k^2 + 2ik\sin\xi
+\cos\xi-\mu_R}\,.~~
\end{eqnarray}
The determinant can be seen as polynomial in $e^{2ikL}$ of order
$2M$ whose roots are functions of $k$ of the form
\begin{eqnarray}
\label{root_periodic}
X_p^\pm=\frac{1}{2}\left[{\omega^*}^p\;T^{+}+\omega^p\;T^{-} \pm
\sqrt{({\omega^*}^p\;T^{+}-\omega^p\;T^{-})^2+4R^+R^-}\right]~~,~~p=0,..,M-1\,,
\end{eqnarray}
where $\omega=e^\frac{2i\pi}{M}$ is the M$^{th}$ root of unity.
For M=1, we recover the result given in \cite{bart}. To get the
spectrum, one has to solve in $k$ the following equations
\begin{eqnarray}
\label{bethe_periodic}
e^{2ikL}=X_p^\pm~~,~~p=0,..,M-1\,.
\end{eqnarray}
Let us introduce the shift operator $\widehat{P} f(x)=f(x+2L)$. In
the case under consideration, this operator commutes with the
Hamiltonian and can therefore be diagonalized in the same basis.
When acting on $\Phi$, this operator is represented by the
following $2M\times 2M$ matrix
\begin{eqnarray}
\widehat{P}= \left(\begin{array}{cccccc}
 &&1 \\
 &&&1\\
 &&&&\ddots\\
 &&&&&1\\
 1 &  \\
 &1
 \end{array}\right)\;.
\end{eqnarray}
It is interesting to note that the eigenfunctions (\ref{ansatz})
constrained by relation (\ref{bethe_periodic}) for a given $p$ are
actually common eigenfunctions of $H$ and $\widehat{P}$ as one can
see from
\begin{eqnarray}
\widehat{P}~\Phi(x)=\omega^p~\Phi(x)\;.
\end{eqnarray}
This relation allows us to give a simple physical interpretation
for $p$. Indeed, $p$ is the usual crystal momentum\footnote{We
choose the units so that $\hbar=1$} carried by the particle. In
the usual approach, it labels the Bloch's functions and we recover
here by our different approach that it is indeed a good quantum
number for a very general periodic potential. We also remark that
equation (\ref{bethe_periodic}) is M-periodic in terms of $p$.
This statement is equivalent to the standard restriction of the
range of the crystal momentum to the first Brillouin zone.

\subsection{Kronig-Penney Model}

\begin{figure}[htp]
\begin{minipage}[t]{8cm}
\epsfig{file=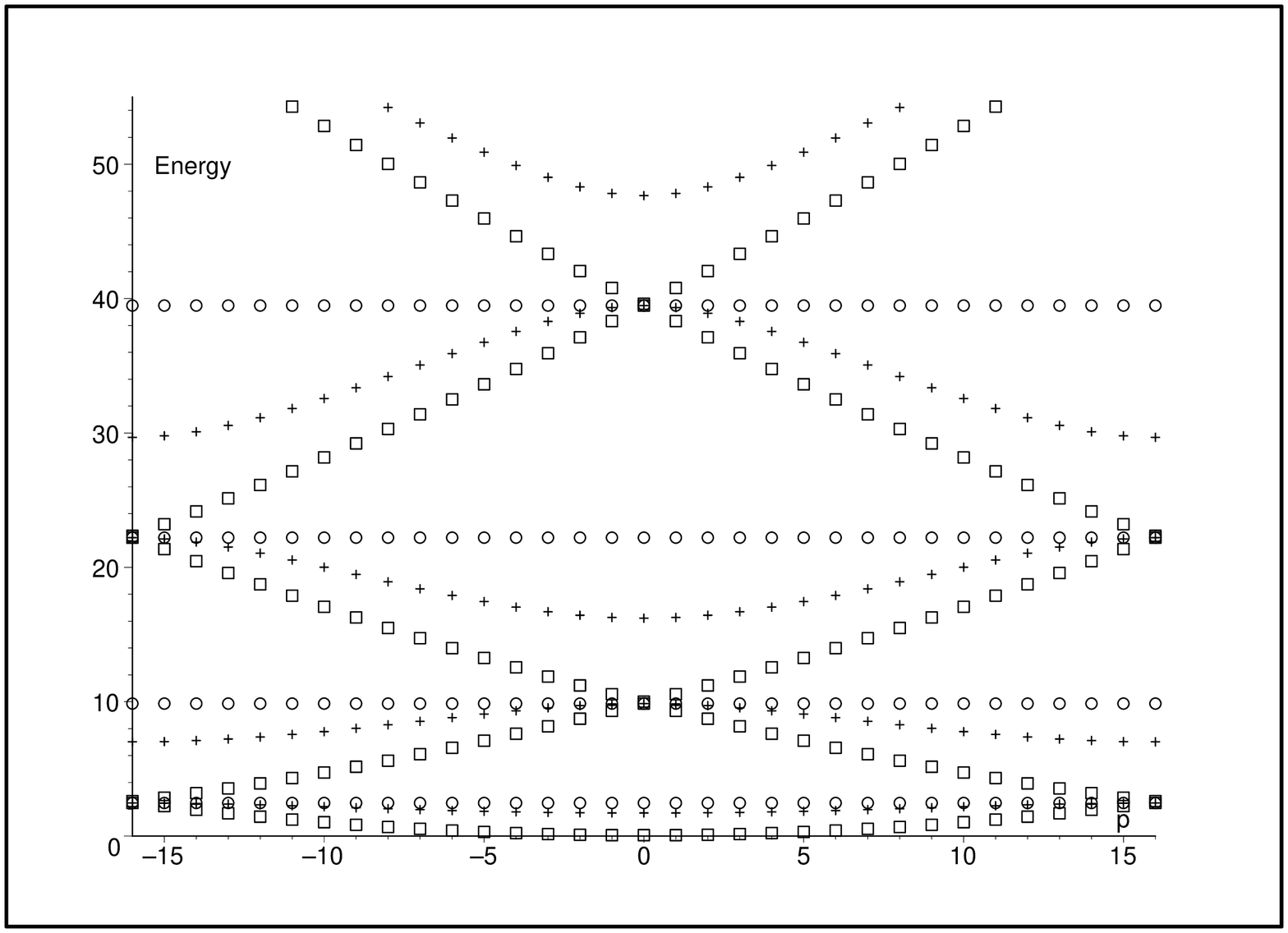,width=6cm,angle=-90}
\caption{\label{fig-delta}Lowest positive energy levels in terms of
$p$ for $\xi=0$(circle); 0.2(cross); $\pi/2$(box)}
\end{minipage}
\qquad
\begin{minipage}[t]{8cm}
\epsfig{file=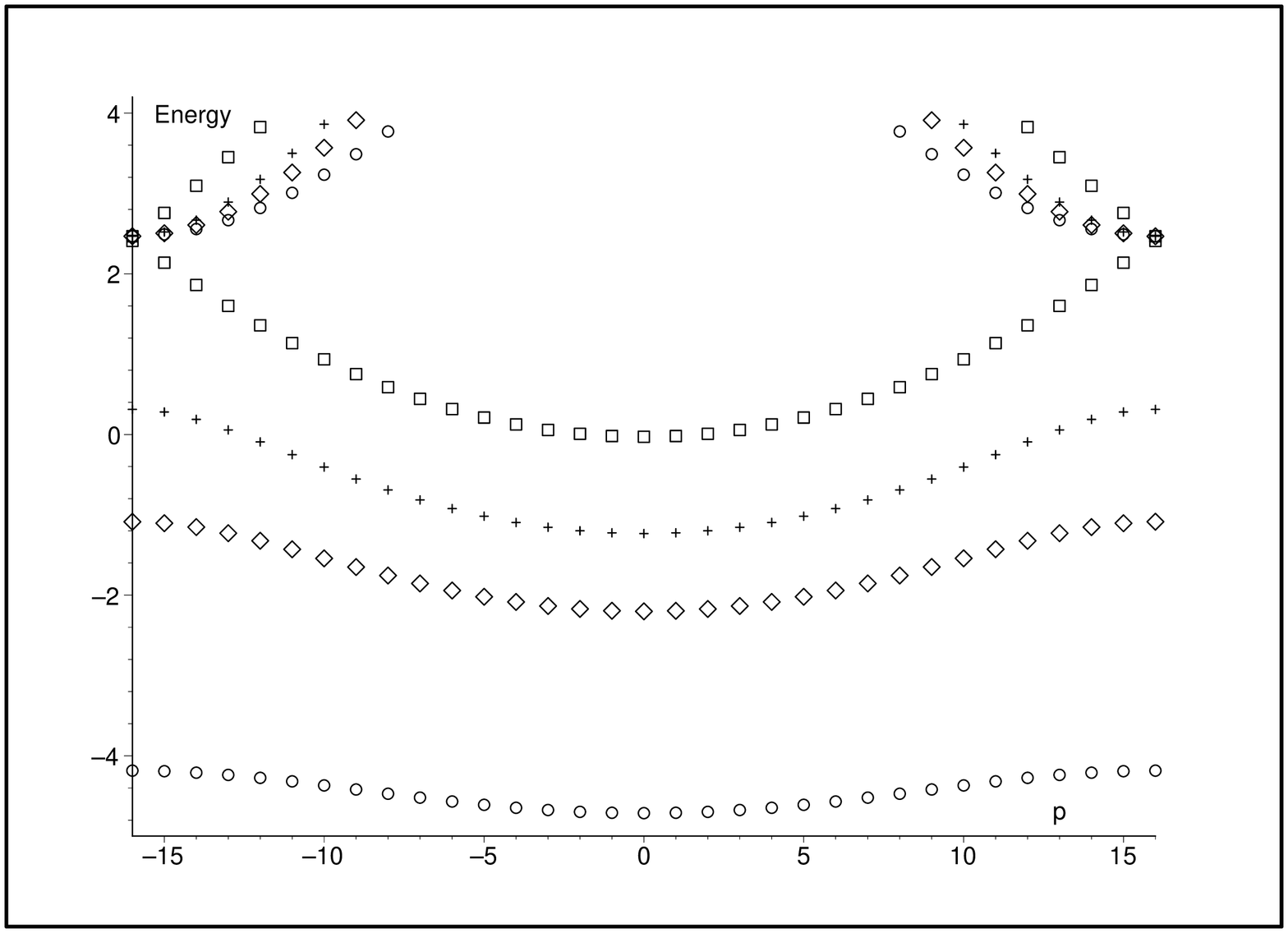,width=6cm,angle=-90}
\caption{\label{fig-deltalinked} Bound states and lowest positive
energy levels in terms of $p$ for $\xi=\pi/2$(box); 2.3(cross);
2.5(diamond); 2.7(circle).}
\end{minipage}
\end{figure}

We now turn to the Kronig-Penney model to show that the equations
(\ref{bethe_periodic}) consistently reproduce the standard
equations of this model. In this case, all the $v_j$'s are given
by a $\delta$ potential with the same coupling constant $c$. This
is obtained from (\ref{perio_condition1}) by taking
\begin{eqnarray}
\label{UKP} U=-e^{i\xi} \left(
\begin{array}{cc}
  \cos(\xi) & i\sin(\xi) \\
  i\sin(\xi) & \cos(\xi) \\
\end{array}
\right)~~ ,~~\xi\in[0,\pi) \qmbox{and} \tan(\xi)=\frac{2}{c}\;.
\end{eqnarray}
Therefore, we get
\begin{eqnarray}
T^+=T^-=\frac{ik\tan(\xi)}{1+ik\tan(\xi)}\qmbox{and}
R^+=R^-=\frac{-1}{1+ik\tan(\xi)}\;.
\end{eqnarray}
After some algebra, we see that the equations
(\ref{bethe_periodic}) become, for p=0,..,M-1 and real $k$
\begin{eqnarray}
\label{bethe_periodic_real}
\cos(2kL)+\frac{1}{k\tan(\xi)}\sin(2kL)=\cos\left(\frac{2\pi
p}{M}\right)\;.
\end{eqnarray}
These are the equations obtained by using Bloch's theorem (see for
example \cite{kittel}). They give the well-known plots that we
show in Figure \ref{fig-delta}\footnote{All the numerical
resolutions and plots are realized with Maple}. The plots
represent the energy spectrum in terms of the crystal momentum $p$
(restricted to the first Brillouin zone)  for various values of
$\xi$ (here $M=32$ and $L=1$). For $\xi=0$ (circle on Figure
\ref{fig-delta}), the coupling constant, $c$, tends to $+\infty$
and the energy does not depend on $p$. The states are localized in
the regions $R^-_{j-1}\cup R^+_{j}$, $j=2,\ldots,M$ which are
separated by purely reflecting walls (there is no transmission).
The choice $\xi=\pi/2$ (box on Figure \ref{fig-delta}) correspond
to another particular case where the coupling constant vanishes.
The states are delocalized and this is just the free particle
model. The energy varies like the square of $p$. For an
intermediate case, here $\xi=0.2$ (cross on Figure
\ref{fig-delta}), the well-known energy band structure appears.

Let us stress that in order to find equation
(\ref{bethe_periodic_real}), we assumed $k$ real. However, to
study the possible bound states we have to look for purely
imaginary solutions of (\ref{bethe_periodic}) of the type
$k=ik_I$, $k_I\in \RR$. In this case, we are led to solve the
following equations, for $p=0,..,M-1$
\begin{eqnarray}
\label{bethe_periodic_imaginar}
(1+k_I\tan\xi)e^{2k_IL}=k_I\tan\xi\cos\left(\frac{2\pi
p}{M}\right) \pm\sqrt{1-k_I^2\tan^2\xi\sin^2\left(\frac{2\pi
p}{M}\right)} \;.
\end{eqnarray}
Figure \ref{fig-deltalinked} shows the solutions of this equation
(for $\xi$=2.3(cross); 2.5(diamond); 2.7(circle)) which gives
negative energy levels. We represent also the lowest positive
energy levels solution of (\ref{bethe_periodic_real}). The bound
states appear only when $\xi>\pi/2$ (i.e. $c<0$). For
$\xi\rightarrow \pi^-$, the coupling constant tends to $-\infty$
and the states are localized around the points $x_j^0$.

\subsection{Asymmetric potential \label{sec_asym}}

\begin{figure}[htp]
\begin{minipage}[t]{8cm}
\epsfig{file=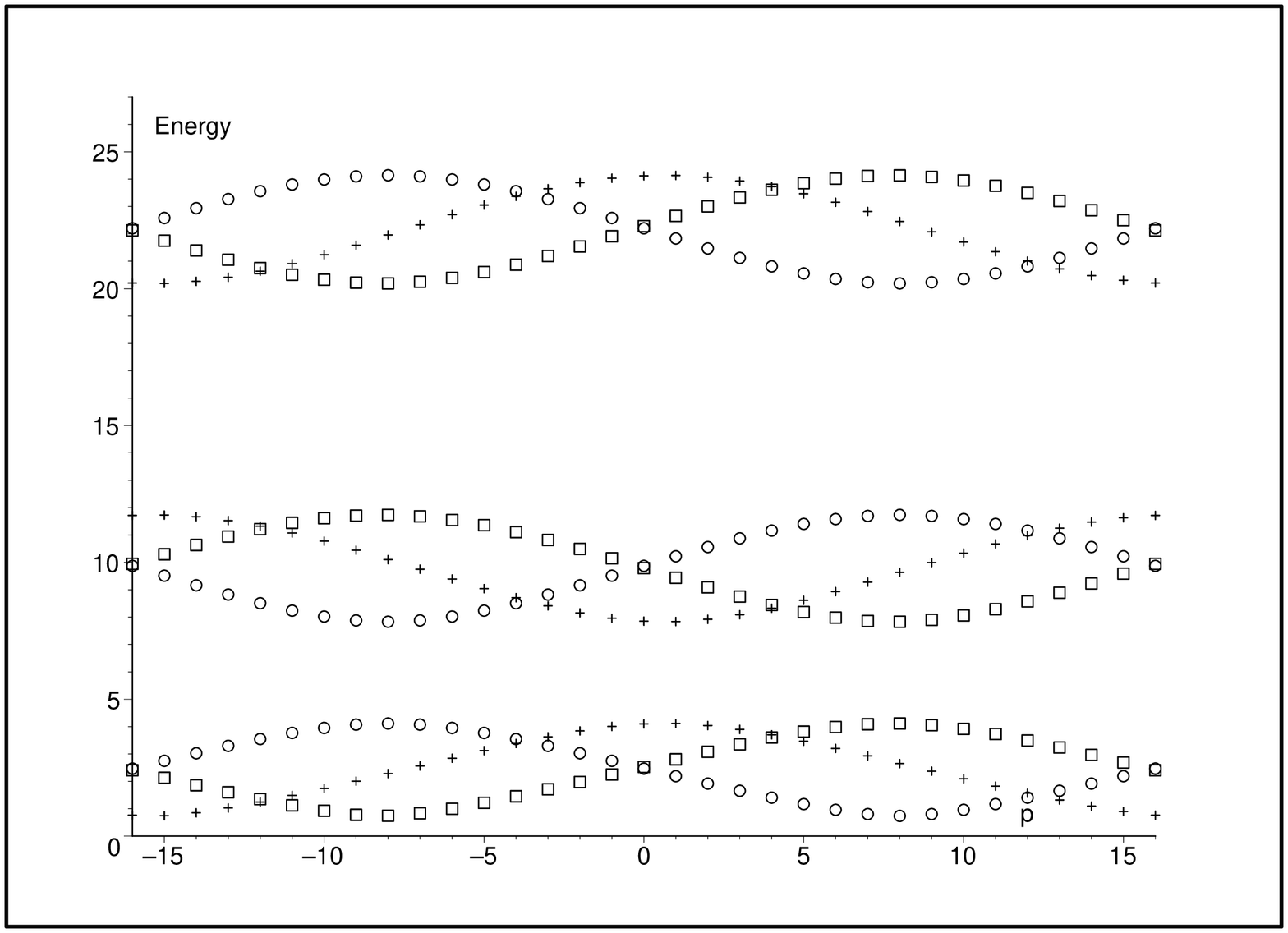,width=6cm,angle=-90}
\caption{\label{fig-asye}Lowest positive energy levels in terms of
$p$ ($\epsilon=-1$) for $\alpha=0$(circle); $\pi/2$(cross);
$\pi$(box)}
\end{minipage}
\qquad
\begin{minipage}[t]{8cm}
\epsfig{file=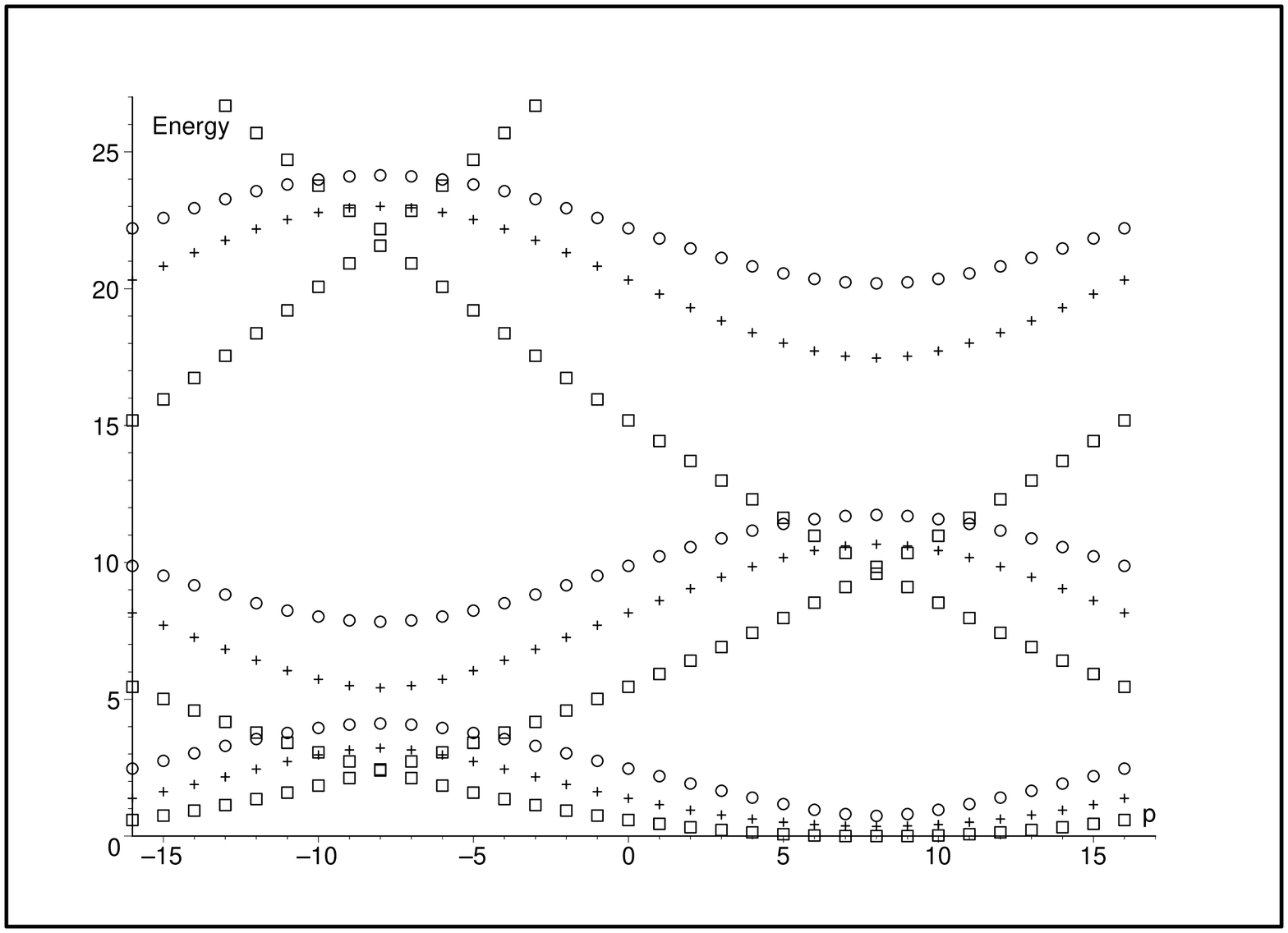,width=6cm,angle=-90}
\caption{\label{fig-asyee} Lowest positive energy levels in terms
of $p$ ($\epsilon=+1$) for $\alpha=0$(circle); $\pi/4$(cross);
$\pi/2$(box).}
\end{minipage}
\end{figure}

In this section, we study the periodic potential characterized by
\begin{equation}
\label{Uasym} U= \left(
\begin{array}{cc}
  0 & e^{i\alpha} \\
  -e^{\epsilon i \alpha} & 0 \\
\end{array}\right)
\qmbox{where}\epsilon=-1 \qmbox{(resp. $\epsilon=1$)}\,.
\end{equation}
This particular form of $U$ for $\epsilon=-1$ (resp.
$\epsilon=+1$) is obtained from (\ref{perio_condition1}) by
setting $\mu=0,~\xi=0$ and $\nu=e^{i\alpha}$ (resp. $\mu=0,~\nu=1$
and $\xi=\alpha$). In this case, one gets
\begin{eqnarray}
&&R^+=\frac{k^2-1}{k^2+1}=R^-\qmbox{and} T^\pm=\frac{\mp2k
e^{\pm i\alpha}}{k^2+1}\\
\mbox{(resp.}&& R^+=\frac{k^2-1}{k^2+2ik\tan\alpha+1}=R^-\qmbox{and}
T^\pm=\frac{\mp2k}{\cos\alpha(k^2+2ik\tan\alpha+1)} )
\end{eqnarray}
For $k\in\RR$, equation (\ref{bethe_periodic}) reduces to
\begin{eqnarray}
\label{be_asym} &&(k^2+1)\sin(2kL)=2k\sin\left(\frac{2\pi
p}{M}-\alpha\right)
\\
\label{be_asyp}
(\mbox{resp.}~&&(k^2+1)\cos\alpha\;\sin(2kL)+2k\sin\alpha\;\cos(2kL)
=2k\sin\left(\frac{2\pi p}{M}\right)~)\;.
\end{eqnarray}
Figure \ref{fig-asye} (resp. Figure \ref{fig-asyee}) shows the
behavior of the solutions of equation (\ref{be_asym}) (resp.
equation (\ref{be_asyp})).
Let us remark that, for a generic $\alpha$, these potentials break
the symmetry of the spectrum under the transformation $p\to -p$.
We also remark that the behaviors of the two spectra are
completely different. For $\epsilon=-1$, the gaps between the
energy bands are independent of the parameters $\alpha$ whereas,
for $\epsilon=+1$, they decrease when $\alpha$ increases.

\section{Dirichlet or Neumann conditions \label{box}}

In this section, we want to study the behavior of the energy band
structure when we modify the boundary conditions. To illustrate
this on a simple example we take the same potential as in
(\ref{perio_condition1}) and simply change $U_q$ in
(\ref{perio_condition2}) to
\begin{equation}
U_q=\varepsilon \left(
\begin{array}{cc}
  1 & 0 \\
  0 & 1 \\
\end{array}\right)
\qmbox{where}\varepsilon=-1 \qmbox{(resp.}\varepsilon=+1)\;.
\end{equation}
This implements the Dirichlet (resp. Neumann) boundary condition:
\begin{equation}
\phi(0)=0=\phi(L) \qmbox{(resp.} \phi'(0)=0=\phi'(L))\,,
\end{equation}
and breaks the periodicity of the model. This constitutes a basic
example where our method still holds while Bloch's theorem breaks
down. In this case the Bethe equations (\ref{bethe}) are
equivalent to
\begin{eqnarray}
\label{bethe_box}
\left|\begin{array}{ccccccccc}
R^+-\varepsilon e^{2ikL}&T^+&&&&&&\\
T^-&R^-&-e^{2ikL}&&\\
&-e^{2ikL}&R^+&T^+\\
&&T^-&R^-&-e^{2ikL}&&\\
&&&&\ddots\\
&&&&T^-&R^-&-e^{2ikL}&\\
&&&&&-e^{2ikL}&R^+&T^+\\
&&&&&&T^-&R^--\varepsilon e^{2ikL}
\end{array}\right|=0\,.
\end{eqnarray}
Again, this determinant is a polynomial of order $2M$ in
$e^{2ikL}$ whose roots read
\begin{eqnarray}
\label{root_box}
Y_q^\pm&=&\sqrt{T^+T^-}\cos\left(\frac{\pi
q}{M}\right)\pm \sqrt{R^+R^--T^+T^-\sin^2\left(\frac{\pi
q}{M}\right)}~~,~~1\leqslant q \leqslant M-1\,,\\
Y^\pm&=&\frac{1}{2}\left[-\varepsilon (R^++R^-)
\pm\sqrt{(R^+-R^-)^2+4T^+T^-}\right]\,.
\end{eqnarray}
The Bethe equations governing the spectrum then read
\begin{eqnarray}
\label{bethe_box1}
e^{2ikL}&=&Y_q^\pm~~,~~1\leqslant q \leqslant M-1\,,\\
\label{bethe_box2} e^{2ikL}&=&Y^\pm\,.
\end{eqnarray}
Let us make a few remarks before showing the influence of the
boundary on the energy spectrum for a particular example. First,
the parameter $q$ cannot be interpreted as the crystal momentum
any more (the shift operator $\widehat P$ does not commute with
the Hamiltonian). It does not label all the energy states since it
is absent from equation (\ref{bethe_box2}). On the other hand,
only equation (\ref{bethe_box2}) depends on the parameter
$\varepsilon$ characterizing the type of boundary conditions under
consideration. We also stress that, although they look different,
the roots in (\ref{root_periodic}) and those in (\ref{root_box})
can be written in the same form provided one replaces $M$ by $M/2$
in (\ref{root_box}) and relabels $q$ appropriately. From all this,
we conclude that equations (\ref{bethe_box2}) give the energy
states arising from the presence of the boundary while the rest of
the spectrum will be identical to that of the periodic case. In
particular, the boundary effects become negligible in the
thermodynamic limit.

\textbf{Example}

We take once again the matrix $U$ given by (\ref{UKP}) to define
the potential at each site but now, we impose Neumann boundary
conditions. The Bethe equations (\ref{bethe_box1}) and
(\ref{bethe_box2}) become respectively (for $k$ real)
\begin{eqnarray}
\label{bethe_box1KP}
&&\cos(2kL)+\frac{1}{k\tan\xi}\sin(2kL)=\cos\left(\frac{\pi
q}{M}\right)~,\qmbox{for} 1\leqslant q \leqslant
M-1\\
\label{bethe_box2KP} &&\cos(2kL)-k\tan(\xi)\sin(2kL)=1\;.
\end{eqnarray}
On Figure \ref{fig-delta-box} below, we plot solutions of
equations (\ref{bethe_box1KP}) and (\ref{bethe_box2KP}) for $M=16$
and $\xi=0.2$. The solutions of (\ref{bethe_box1KP}) are
represented by circles and those of (\ref{bethe_box2KP}) by boxes
which we displayed at $q=0$ and $q=16$ for convenience. To make
the comparison even easier, we have also represented the energy
band structure of the periodic case for $M=32$ (crosses on Figure
\ref{fig-delta-box}) . We see that for the first and third energy
band, the allowed energy levels coincide with the periodic case as
the solutions of (\ref{bethe_box2KP}) nicely complete those of
(\ref{bethe_box1KP}). However, the effect of the boundary
conditions shows up in the second energy band where two states are
"missing" with respect to the periodic case. Again we recover
graphically that this effect will become more and more negligible
as $M$ becomes larger.

\begin{figure}[htp]
\begin{center}
\epsfig{file=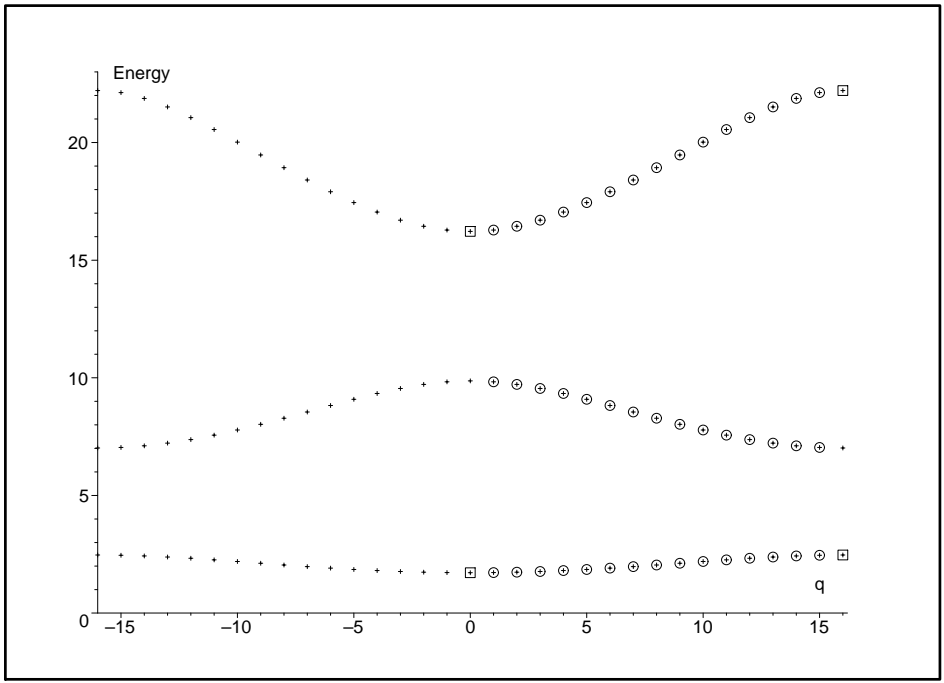,width=9cm,angle=-90}
\caption{\label{fig-delta-box}Lowest positive energy levels in terms
of $q$ for $\xi=0.2$}
\end{center}
\end{figure}

\section{Impurities \label{impurity}}

So far, we have considered homogeneous potentials in the sense
that the potential was the same at each site. As illustrated
above, even with non-periodic boundary conditions, this does not
deviate drastically from the periodic case as only local boundary
effects arise.

In this section, we want to explore very different situations
which cannot be approached by perturbing the periodic case and
therefore, show the advantages of our approach. Such situations
occur when one modifies the potential at only one or several
sites. Thus, these sites appear as what we call impurities. These
models can be useful for describing localized defects in materials
with one dimensional behaviour (nanowires, etc.)

We illustrate this in the following by considering one such
impurity and then a subset of equally spaced identical impurities
in a given periodic potential.

\subsection{One impurity}

We restrict ourselves to the case of periodic boundary conditions
and we imagine that the impurity is sitting at site $1$ with a
potential $v_1$ given as in (\ref{litv}) while all the other sites
have the same potential $v_2=\ldots=v_M=v$. This model is obtained
in our language by taking
\begin{eqnarray}
\label{imp_condition1}
U_1=e^{i\xi_1}\left(\begin{array}{cc}
                        \mu_1 & \nu_1 \\
                         -\nu_1^* & \mu_1^* \\
                       \end{array}\right)&,&
U_2=\dots=U_M=U=e^{i\xi}\left(\begin{array}{cc}
                        \mu & \nu \\
                         -\nu^* & \mu^* \\
                       \end{array}\right)
~\mbox{and}~U_q= \left(
\begin{array}{cc}
  0 & 1 \\
  1 & 0 \\
\end{array}\right)
\end{eqnarray}
where $U_1$ and $U$ characterize respectively $v_1$ and $v$. In
this case, the Bethe equations (\ref{bethe}) reduce to
\begin{eqnarray}
\label{bethedet_imp} \left|\begin{array}{ccccccccc}
R_1^+&T_1^+&&&&&&-e^{2ikL}\\
T_1^-&R_1^-&-e^{2ikL}&&\\
&-e^{2ikL}&R^+&T^+\\
&&T^-&R^-&-e^{2ikL}&&\\
&&&&\ddots\\
&&&&T^-&R^-&-e^{2ikL}&\\
&&&&&-e^{2ikL}&R^+&T^+\\
-e^{2ikL}&&&&&&T^-&R^-
\end{array}\right|=0
\end{eqnarray}
where $R_1^+$, $T_1^+$, $T_1^-$ and $R_1^-$ are given by relations
(\ref{RTRT1}) and (\ref{RTRT2}) substituting $\xi$, $\mu$ and $\nu$
by $\xi_1$, $\mu_1$ and $\nu_1$, respectively.

To simplify the expression of these Bethe equations, we consider
the case of symmetric potentials \ie $R^+=R^-=R$, $T^+=T^-=T$,
$R^+_1=R^-_1=R_1$ and $T^+_1=T^-_1=T_1$. In this particular case,
the vanishing of determinant (\ref{bethedet_imp}) is equivalent to
the following equations
\begin{eqnarray}
\label{bethe_imp} 2e^{2ikL}=\left(y+\frac{1}{y}\right)T \pm
\sqrt{\left(y-\frac{1}{y}\right)^2T^2+4R^2}
\end{eqnarray}
where $y$ is any root of one of the following polynomials
\begin{eqnarray}
&&\hspace{-1cm}T^2(T_1\mp R_1)(y^{2M}+1) +T(R^2-T^2-R_1^2-T_1^2\pm
2T_1R)y^M\nonumber\\
\label{poly-y} &&\hspace{4cm}+ R\Big(2T_1(R-R_1)\pm(R-R_1)^2\pm
T_1^2\mp T^2\Big)\sum_{j=1}^{M-1}y^{2j}\;.
\end{eqnarray}
Note that we consistently recover the periodic case when $R_1=R$
and $T_1=T$ since the polynomials (\ref{poly-y}) simply reduce to $(y^M-1)^2$.\\

\textbf{Example}

We imagine that both the impurity and the bulk potential are given
by a $\delta$ potential but with different coupling constants (\ie
the matrices $U$ and $U_1$ are given by (\ref{UKP}) with parameter
$\xi$ and $\xi_1$, respectively). We study the behavior of the
spectrum when the coupling constant of the impurity is varied
while $\xi$ is kept fixed (we choose here $\xi=0.2$).

\begin{figure}[htp]
\begin{minipage}[t]{8cm}
\epsfig{file=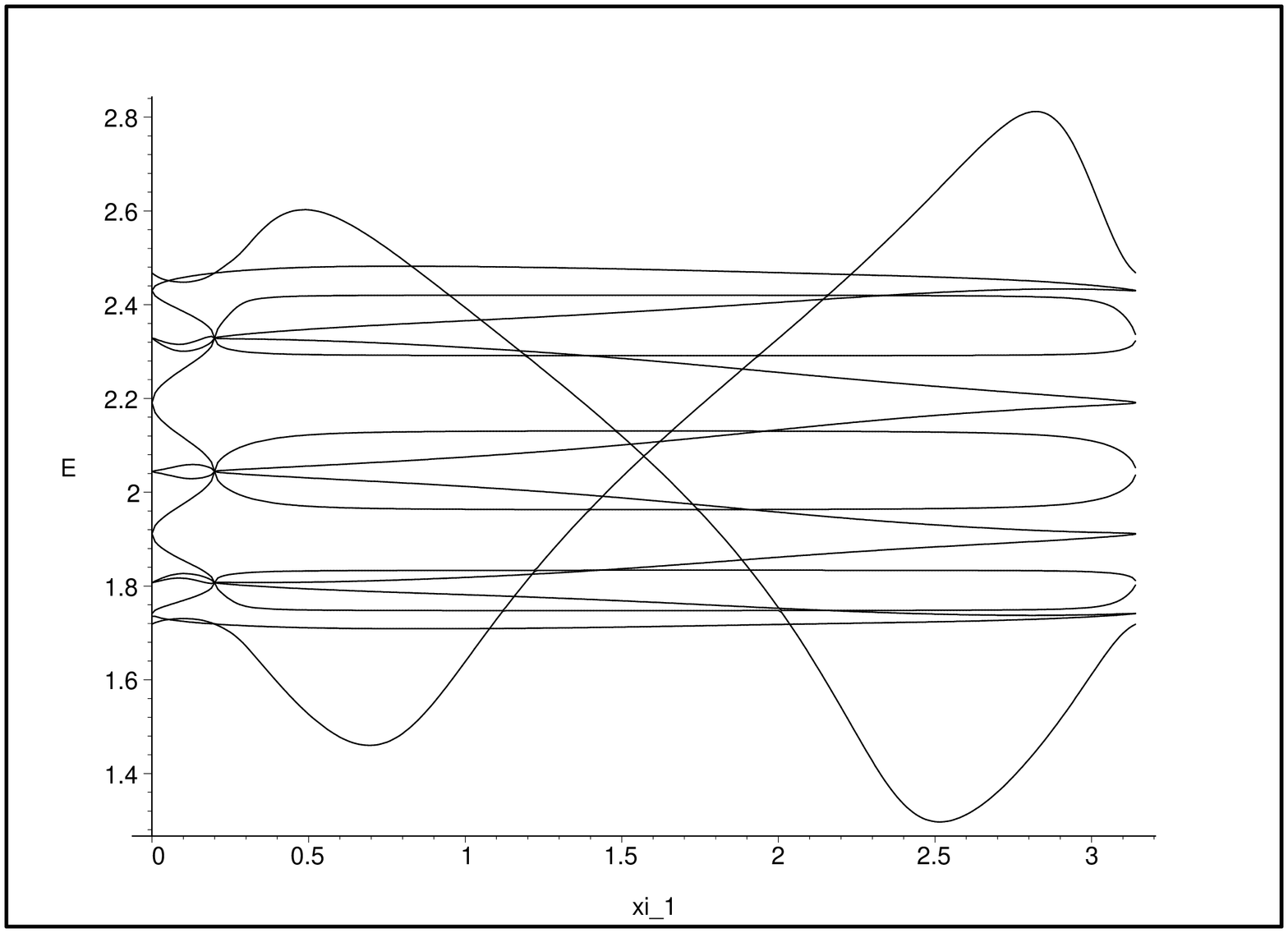,width=6cm,angle=-90}
\caption{\label{fig-delta-imp}Energy levels of the first band in
terms of $\xi_1$ for $\xi=0.2$}
\end{minipage}
\qquad
\begin{minipage}[t]{8cm}
\epsfig{file=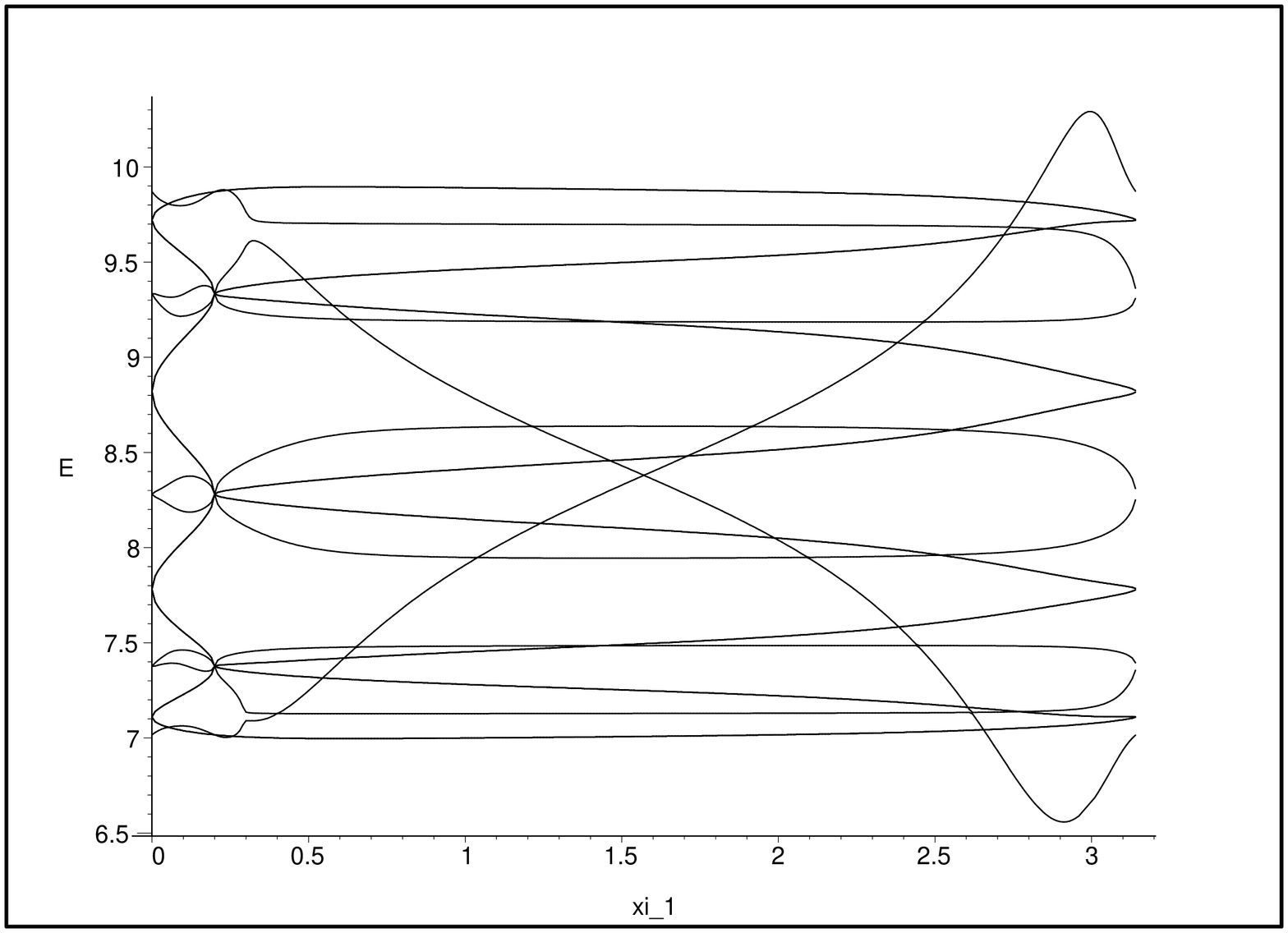,width=6cm,angle=-90}
\caption{\label{fig-delta-imp1}Energy levels of the second band in
terms of $\xi_1$ for $\xi=0.2$}
\end{minipage}
\end{figure}

Figures \ref{fig-delta-imp} and \ref{fig-delta-imp1} show the
behavior of the energy levels of the first and second positive
energy bands as functions of $\xi_1$ for $M=8$. In the case where
$\xi_1=\xi=0.2$, we remark a degeneracy as the impurity becomes
identical to the bulk potential. The corresponding energy levels
coincide with those represented by crosses on Figure
\ref{fig-delta}. Generally speaking, the effect of the impurity is
far from being trivial. We see that in each band, two energy
levels are severely modified depending on the strength of the
impurity. This can have strong consequences on the corresponding
system (if we think of conducting or insulating materials for
instance).

\subsection{$N$ identical impurities}

Here, we introduce periodically $N$ identical impurities ($N\in
\NN$) in a given periodic potential. From now on, let $M$ be a
multiple of $N$ i.e. $M=M_1 N$ with $M_1>1$. The impurities have
potential $v_1$ and the bulk potential is $v$. In this case, the
matrices $U_j$ take the following particular form
\begin{eqnarray}
\label{imps_condition1}
&&U_1=U_{M_1+1}=\dots=U_{(N-1)M_1+1}=e^{i\xi_1}\left(\begin{array}{cc}
                        \mu_1 & \nu_1 \\
                         -\nu_1^* & \mu_1^* \\
                       \end{array}\right)\\
 \label{imps_condition2}
&&U_2=\dots=U_{M_1}=U_{M_1+2}=\dots=U_{(N-1)M_1+2}=\dots=U_M=
e^{i\xi}\left(\begin{array}{cc}
                        \mu & \nu \\
                         -\nu^* & \mu^* \\
                       \end{array}\right)
\end{eqnarray}
As before, we restrict ourselves to the case of symmetric
potentials for simplicity ($R^+=R^-=R$, $T^+=T^-=T$,
$R^+_1=R^-_1=R_1$ and $T^+_1=T^-_1=T_1$). The Bethe equations take
the form
\begin{eqnarray}
\label{bethe_imps} 2e^{2ikL}=T\left(y+\frac{1}{y}\right) \pm
\sqrt{T^2\left(y-\frac{1}{y}\right)^2+4R^2}\,.
\end{eqnarray}
Again, $y$ is any root of one of the following polynomials, for
$0\leqslant q \leqslant N-1$
\begin{eqnarray}
\label{poly-yw}
4\gamma^{2q}\;P_q^+(y)P_q^-(y)+y^{2M_1}(\gamma^q-1)^2(\gamma^q+1)^2
T^2(R^2-T^2+T_1^2-R_1^2)^2\,,
\end{eqnarray}
where we have introduced
\begin{eqnarray}
&&\hspace{-1cm}P_q^\pm(y)=T^2(T_1\mp R_1)(y^{2M_1}+1)
+\frac{1}{2}\left(\gamma^q+\frac{1}{\gamma^q}\right)
T(R^2-T^2-R_1^2-T_1^2\pm 2T_1R)y^{M_1}\nonumber\\
&&\hspace{4.8cm}+ R\Big(2T_1(R-R_1)\pm(R-R_1)^2\pm T_1^2\mp
T^2\Big)\sum_{j=1}^{M_1-1}y^{2j}\,,
\end{eqnarray}
and $\gamma=e^{\frac{2i\pi}{N}}$ is the $N^{th}$ root of unity.
The presence of $\gamma$ accounts for the inclusion of $N$
identical impurities. In the case $N=1$, one has $\gamma=1$ and
the set of coupled equations (\ref{bethe_imps}),(\ref{poly-yw})
reduces to (\ref{bethe_imp}),(\ref{poly-y}).

\section{$\cN$ bosons with contact interaction}\label{interaction}

In this section, we extend the general method of Section
\ref{gen_pb} to $\cN$ bosons with $\delta$-interaction in presence
of the potential defined in (\ref{ham}),(\ref{litv}). We restrict
ourselves to the case of bosons for the sake of clarity but a
treatment analogous to that in \cite{Y,bart}, where the statistics
is not given {\it a priori}, is possible. Our aim is to derive the
corresponding Bethe ansatz equations which can be used then as
above to get information on the energy spectrum of the model.
However, we will not go into the details of exploring particular
cases as the essential features of our approach have already been
discussed in the previous sections. Instead, we trust that the
interested people will adapt the general equations to their
specific needs in order to study the influence of interactions in
particular models with point potentials.

We first need some definitions and notations. Let $\mS_{\cN}$ and
$\mW_{\cN}$ denote the permutation group and the Weyl group
associated to the Lie algebra $B_\cN$ respectively. The group
$\mS_{\cN}$ consists of $\cN$ generators: the identity $Id$ and
$\cN-1$ elements $T_1,\ldots,T_{\cN-1}$ satisfying
\begin{eqnarray}
\label{TT} &&T_j\,T_j=Id
\qmbox{,}T_{j}T_{\ell}=T_{\ell}T_{j} \qmbox{for} |j-\ell|>1\qmbox{,}\\
\label{TTT} &&T_{j}T_{j+1}T_{j}=T_{j+1}T_{j}T_{j+1}\qmbox{.}
\end{eqnarray}
In particular, the last relation gives rise to the famous
Yang-Baxter equation \cite{Y,baxter}. For convenience, we denote a
general transposition of $\mS_{\cN}$ by $T_{ij}$, $i< j$, given by
\begin{equation}
\label{generalT} T_{ij}=T_{j-1}\ldots T_{i+1}T_iT_{i+1}\ldots
T_{j-1}
\end{equation}
The Weyl group $\mW_{\cN}$ contains $2^{\cN}\cN!$ elements
generated by $Id$, $T_1,\ldots,T_{\cN-1}$ and $R_1$ satisfying
(\ref{TT}), (\ref{TTT}) and
\begin{eqnarray}
&&R_1\,R_1=Id\qmbox{,}\\
&&R_1T_{1}R_1T_{1}=T_{1}R_1T_{1}R_1\qmbox{,}\\
&&R_1T_{j}=T_{j}R_1\qmbox{for} j>1\, .
\end{eqnarray}
Let us define also $R_j$, $j=2,\ldots,N$ as
\begin{equation}
R_j=T_{j-1}\ldots T_1R_1 T_1\ldots T_{j-1}\;.
\end{equation}
Let $x_1,\dots,x_{\cN}$ be the positions of the particles. The
natural generalization of (\ref{Phi}) leads to represent the
wavefunction as follows
\begin{eqnarray}
\phi(x_1,\dots,x_\cN)&=&
\phi^{\epsilon_1,\dots,\epsilon_{\cN}}_{j_1,\dots,j_{\cN}}(x_1,\dots,x_\cN)
\end{eqnarray}
in the regions, for $1\leqslant k \leqslant \cN$,
\begin{eqnarray}
\left(M-2j_k+\frac{\epsilon_k+1}{2}\right)L<x_k
<\left(M-2j_k+\frac{\epsilon_k+3}{2}\right)L\,,
\end{eqnarray}
and for $x_1,\dots,x_{\cN}$ two by two different. Then, one
defines the following $(2M)^{\cN}$-component vector
\begin{eqnarray}
 \Phi(x_1,\dots,x_\cN)=
\mathop{\sum_{1\leqslant j_1,\dots,j_{\cN}\leqslant
M}}_{\epsilon_1,\dots,\epsilon_{\cN}=\pm}
\phi^{\epsilon_1,\dots,\epsilon_{\cN}}_{j_1,\dots,j_{\cN}}
\left(f^{\epsilon_1}_{j_1}(x_1),\dots,f^{\epsilon_{\cN}}_{j_{\cN}}(x_{\cN})\right)
e_{j_1}\otimes\hat{e}_{\epsilon_1}\otimes \dots\otimes
e_{j_{\cN}}\otimes\hat{e}_{\epsilon_{\cN}}
\end{eqnarray}
where, for $1\leqslant k \leqslant \cN$,
\begin{eqnarray}
f^{\pm}_k(x)=\pm(x+2kL-L-ML)-L+ML\;.
\end{eqnarray}
As before, the advantage of the vector $\Phi(x_1,\dots,x_\cN)$ is
that it contains all the information on the wavefunction for
$x_1,\dots,x_{\cN}$ running only in the interval $]ML-L,ML[$ of
length $L$. This allows to impose all the conditions for
interactions between particles and with the external potential in
a very compact form.

Given a tensor product of spaces, $(\CC^{2M})^{\otimes \cN}$, we
define the action of a matrix $A\in End(\CC^{2M})$ on the $k^{th}$
space by
\begin{equation}
\label{tensor} A^{[k]}=\underbrace{\II \otimes \dots \otimes
\II}_{k-1} \otimes A \otimes \underbrace{\II \otimes \dots \otimes
\II}_{\cN-k}\;.
\end{equation}
\newpage
We are now in position to write all the conditions the
wavefunction of the problem has to satisfy:
\begin{itemize}

\item Shr\"odinger equation: for $x_i\in]ML-L,ML[$ and
$x_1,\dots,x_{\cN}$ two by two different
\begin{eqnarray}
-\sum_{k=1}^\cN \partial^2_{x_k}\Phi(x_1,\dots,
x_{\cN})=E\Phi(x_1,\dots, x_{\cN})\,.
\end{eqnarray}

\item External point potential: for $1\leqslant k \leqslant \cN$
\begin{eqnarray}
&&\left[(\cU^+-\II)^{[k]}+i(\cU^++\II)^{[k]}~\partial_{x_k}\right]
\Phi(x_1,\dots, x_{\cN})=0\qmbox{for} x_k\rightarrow ML-L\,,\\
&&\left[(\cU^--\II)^{[k]}+i(\cU^-+\II)^{[k]}~\partial_{x_k}\right]
\Phi(x_1,\dots, x_{\cN})=0\qmbox{for} x_k\rightarrow ML\,.
\end{eqnarray}

\item Interactions between particles: for $Q\in \mS_{\cN}$ and
$1\leqslant i \leqslant \cN-1$,
\begin{eqnarray}
\label{bulk_condition1} &&\hspace{-1cm}\Phi(x_1,\dots,
x_N)|_{~x_{Qi}=x_{Q(i+1)}^+}
=\widetilde{Q}~\widetilde{T_{i}}~\widetilde{Q}^{-1}~\Phi(x_1,\dots, x_N)|_{~x_{Qi}=x_{Q(i+1)}^-}\,,\\
&&\hspace{-1cm}(\partial_{x_{Qi}}-\partial_{x_{Q(i+1)}})~\Phi(x_1,\dots,
x_N)|_{~x_{Qi}=x_{Q(i+1)}^+}\nonumber\\
&&\hspace{+0.5cm}
=\widetilde{Q}~\widetilde{T_{i}}~\widetilde{Q}^{-1}~\left[
(\partial_{x_{Qi}}-\partial_{x_{Q(i+1)}})
+2g\right]\Phi(x_1,\dots,
x_N)|_{~x_{Qi}=x_{Q(i+1)}^-}\,.\label{bulk_condition2}
\end{eqnarray}
where $\widetilde{Q}$ is the usual representation of the element
$Q\in \mS_{\cN}$ on $(\CC^{2M})^{\otimes \cN}$.

\end{itemize}
For bosons, the wavefunction $\phi(x_1,\dots,x_{\cN})$ should be
symmetric under the exchange of any two particles. In terms of
$\Phi$, this reads, for $1 \leqslant i < j \leqslant N$,
\begin{eqnarray}
\label{statistics} \Phi(x_1,\dots,x_i,\dots,x_j,\dots,x_{\cN})=
\widetilde T_{ij}~~ \Phi(x_1,\dots,x_j,\dots,x_i,\dots,x_{\cN})\;.
\end{eqnarray}
Therefore, we can order the particles. In the following, we take
$ML-L<x_1<\dots<x_{\cN}<ML$. The ansatz consists in representing
$\Phi$ by
\begin{eqnarray}
\label{ansatzN} \Phi(x_1,\dots, x_{\cN})= \sum_{P\in~ \mW_{\cN}}
e^{i(k_{P1}x_{1}+\dots+k_{P\cN}x_{{\cN}})}~ \cA_P
\end{eqnarray}
where
\begin{eqnarray}
\cA_P= \mathop{\sum_{1\leqslant j_1,\dots,j_{\cN}\leqslant
M}}_{\epsilon_1,\dots,\epsilon_{\cN}=\pm}
A^{\epsilon_1,\dots,\epsilon_{\cN}}_{P,j_1,\dots,j_{\cN}}
~e_{j_1}\otimes\hat{e}_{\epsilon_1}\otimes \dots\otimes
e_{j_{\cN}}\otimes\hat{e}_{\epsilon_{\cN}}
\end{eqnarray}
and, for any $v=(v_1,\dots,v_{\cN})$,
\begin{eqnarray}
v_{T_i}&=&(v_1,\dots,v_{i-1},v_{i+1},v_i,v_{i+2},\dots,v_{\cN})\\
v_{R_1}&=&(-v_1,v_{2},\dots,v_{\cN})\;.
\end{eqnarray}

We need to determine the $2^{\cN}\cN! (2M)^\cN$ parameters
$A^{\epsilon_1,\dots,\epsilon_{\cN}}_{P,j_1,\dots,j_{\cN}}$ to
find the solution of our problem. Inserting in
(\ref{bound_condition}) and (\ref{bound_condition2}), one gets
\begin{eqnarray}
\label{relZ1} \cA_{PR_1}&=&
\big(Z^+(-k_{P1})\big)^{[1]}~e^{2ik_{P1}(M-1)L}~\cA_{P}\\
\label{relZ2} \cA_{PR_{\cN}}&=&
\big(Z^-(-k_{P\cN})\big)^{[\cN]}~e^{2ik_{P\cN}ML}~\cA_{P}
\end{eqnarray}
where $Z^\pm$ is given by (\ref{defZ}). From relations
(\ref{bulk_condition1}), (\ref{bulk_condition2}), we get for
$1\leqslant j \leqslant \cN-1$
\begin{eqnarray}
\label{relY} \cA_{PT_j}=y(k_{Pj}-k_{P(j+1)})\cA_{P}
\end{eqnarray}
where
\begin{eqnarray}
y(k)=\frac{k-i g}{k+i g}\,.
\end{eqnarray}
Since our construction is based on $\mW_{\cN}$, the Bethe ansatz
is consistent if $Z^+$ and $Y$ satisfy different relations (see
for example \cite{bart}) like the Yang-Baxter equation
\cite{Y,baxter} or the reflection equation \cite{che,skly}. These
relations hold true by direct computation with the explicit form
of the matrices. The non-vanishing of the wavefunction implies the
following constraints, for $1\leqslant j \leqslant \cN$,
\begin{eqnarray}
\det\left[\big(Z^+(-k_j)\big)^{[1]}-\prod_{m \neq j}y(k_{j}+k_{m})
y(k_{j}-k_{m})\big(Z^-(-k_j)\big)^{[\cN]}e^{2ik_{j}L}
 \right]=0 \;.
\end{eqnarray}
These relations are the Bethe equations associated to our problem.
We recover the results of \cite{bart} by setting $M=1$ and
$U_q=\left(\begin{array}{cc}0 & 1\\1& 0\end{array}\right)$. As in
the previous section, to get the spectrum of a given model with
$\delta$ interaction, one chooses the corresponding matrices $U_j$
and $U_q$ and solves these equations in $k_j$. The energy of the
corresponding state is given by $\displaystyle E=\sum_{j=1}^\cN
k_j^2$.

\section*{Conclusions}

In this paper, we have presented a general method to address the
question of the energy spectrum for a large class of
one-dimensional models with equally spaced point potentials. We
also describe how to extend the method to the case where particles
interactions are present. The main results consist in Bethe ansatz
equations which allow a non-perturbative treatment of these
issues.

We have illustrated the method in various typical situations of
interest and have obtained exacts results on the effects of
boundaries and impurities in such models. In view of the
importance of the Kronig-Penney model in solid state physics, we
believe that our method will serve as a useful toolbox to treat
even more realistic situations which now occur in experimental
situations with the advent of the "quantum technology".

\vspace{1cm}

 \textbf{Acknowledgements:} V.C. thanks the UK Engineering and
Physical Sciences Research council for a Research Fellowship. N.C.
is supported by the TMR Network "EUCLID. Integrable models and
applications: from strings to condensed matter", contract number
HPRN-CT-2002-00325.

\end{document}